# A Brief Introduction to POVM Measurement in Quantum Communications

Renzhi Yuan (Email: renzhiyuan90@gmail.com)

## Abstract

This paper gives a brief introduction to Positive-Operator Valued Measure (POVM) of quantum communications. The Projection-Valued Measure (PVM) is first introduced and then the POVM. The relation between POVM and PVM is discussed and an example of POVM in practical measurement is given. This paper provides some insight of POVM measurement for quantum communications.

## 1. Introduction

Quantum measurement extracts the transmitted information from received quantum signals and therefore performs an important role of quantum communications. The simplest quantum measurement is the projection-valued measure (PVM), also called standard measurement or von Neumann measurement, where elementary projectors are usually used [1]. Sometimes the positive-operator valued measure (POVM) is more efficient of obtaining information about the state of a quantum system than a standard measurement [1, 2]. This paper gives a brief introduction to quantum measurement especially the POVM measurement. The Neumark's theorem [3], which declaims that any POVM measurement in a small system can be achieved by a PVM measurement in a larger system, is introduced. And a concrete example of POVM in a practical measurement is given.

## 2. Projection-Valued Measure

A PVM measurement is obtained through a projector system [1], which is defined as a set of operators $\{P_i, i \in M\}$ of Hilbert space $H$, where $M$ is an alphabet set of all possible outcomes of the measurement, if these operators have properties: 1) $P_i$ is Hermitian: $P_i = P_i^\dagger$; 2) $P_i$ is positive semi-definite: $P_i \geq 0$; 3) $P_i$ is idempotent: $P_i^2 = P_i$; 4) $P_i$ is pairwise orthogonal: $P_i P_j = \delta_{ij} = 0$, for $i \neq j$; 5) $\{P_i, i \in M\}$ forms a resolution of the identity on H: $\sum_{i \in M} P_i = I_H$.

The probability of obtaining outcome $i$ for a given state $s = |\psi\rangle$ is specified by
$$p_m(i|\psi) = P(m = i|s = |\psi\rangle) = \langle\psi|P_i|\psi\rangle \tag{1}$$

And the post-measurement state is given by
$$|\psi_{post}^{(i)}\rangle = \frac{P_i|\psi\rangle}{\sqrt{\langle\psi|P_i|\psi\rangle}} \tag{2}$$

For mixed state, specified by the density matrix $\rho$, the probability of obtaining outcome $i$ is given by
$$p_m(i|\rho) = \text{tr}(P_i \rho) \tag{3}$$

where $\text{tr}(\cdot)$ is the trace operation. And the post-measurement state is specified by the following density matrix:
$$\rho_{post}^{(i)} = \frac{P_i \rho P_i}{\text{tr}(P_i \rho)} \tag{4}$$

## 3. Positive-Operator Valued Measure

### 3.1. POVM definition

If we loosen the conditions specified for measurement operators, i.e. omit the idempotent and the orthogonal restrains for the measurement operators, then we can define the Positive-Operator Valued Measure [2]. In a POVM measurement, the measurement operator system $\{Q_i, i \in M\}$ have properties: 1) $P_i$ is Hermitian: $Q_i = Q_i^\dagger$; 2) $P_i$ is positive semi-definite: $Q_i \geq 0$; 3) $\{P_i, i \in M\}$ forms a resolution of the identity on H: $\sum_{i \in M} Q_i = I_H$.

The probability of obtaining outcome $i$ for a given state specified by density matrix $\rho$ is given by
$$p_m(i|\rho) = \text{tr}(Q_i\rho) \tag{5}$$
However, the post-measurement state can not be specified after a POVM measurement. If we wan to specify the post-measurement state of POVM, more information of measurement operators $\{Q_i, i \in M\}$ needs to be given. Thus, define a set of operators $\{A_i, i \in M\}$ and for each $i \in M$:
$$Q_i = A_i^\dagger A_i \tag{6}$$
where $A_i$ is called Kraus operator, and above expression is called a Kraus decomposition of the measurement operator [1]. Obviously, the Kraus decomposition is not unique because for any unitary operator $U$, $UA_i$ is also a Kraus operator. And since $Q_i$ is positive semidefinite, we can always assume that $A_i = \sqrt{Q_i}$. After introducing Kraus operators $\{A_i, i \in M\}$, the post-measurement of POVM is given by
$$\rho_{post}^{(i)} = \frac{A_i \rho A_i}{\text{tr}(A_i^\dagger A_i \rho)} \tag{7}$$

POVM can give more information about the original state. Suppose we want to specify a quantum state is either $|\psi\rangle$ or $|\phi\rangle$, i.e. to discriminate $|\psi\rangle$ and $|\phi\rangle$, where $|\psi\rangle$ and $|\phi\rangle$ are nonorthogonal, which measurement method can be used?

If a PVM measurement is performed, no matter which basis are selected, we have no chance to discriminate these two quantum states certainly, since they are nonorthogonal. For example, if we choose a PVM specified by operators $\{P_0 = |\psi\rangle\langle\psi|, P_1 = |\psi\rangle^\perp\langle\psi|^\perp\}$ and obtain a result 0, still we can not tell whether the original state is $|\psi\rangle$ or not. Because $|\phi\rangle$ also has a probability of $|\langle\psi|\phi\rangle|^2$ to get a result of 0.

However, if a POVM measurement is performed specified by operators $\{Q_0, Q_1, Q_2\}$ [4]:
$$Q_0 = a(I - |\psi\rangle\langle\psi|), Q_1 = a(I - |\phi\rangle\langle\phi|), Q_2 = I - Q_0 - Q_1 \tag{8}$$
where $1/2 \leq a \leq 1$ and $a$ should be carefully chosen as small enough to ensure that $Q_2 > 0$.

Then if the state is $|\psi\rangle$: $p_0 = 0, p_1 = a(1 - |\langle\psi|\phi\rangle|^2), p_2 = 1 - p_1$;

And if the state is $|\phi\rangle$: $p_0 = a(1 - |\langle\psi|\phi\rangle|^2), p_1 = 0, p_2 = 1 - p_0$.

Therefore, if we get result 1, we know the original state is $|\psi\rangle$ definitely, and if we get result 0, we know the original state is $|\phi\rangle$ definitely. Only if we get result 2, we can not tell the original state.

## 3.2. Achieve a POVM measurement using a PVM measurement

In practical measurement, a POVM measurement is achieved by a PVM measurement. In fact, according to the Neumark's theorem, any POVM in a subspace U of Hilbert space $H$ can be achieved by a PVM performed on a larger subspace $\widetilde{U}$ of $H$.

Here gives a brief proof of Neumark's theorem.

Our purpose is to find a set of corresponding PVM operators $\{P_i, i \in M\}$ for a given set of POVM operators $\{Q_i, i \in M\}$. Firstly, we claim that any POVM operators $\{Q_i, i \in M\}$ in subspace $U$ can be achieved by a set of rank 1 POVM operators $\{Q_j, j \in N\}$ in subspace U, where $N \geq M$. And for each $i \in M$:
$$Q_i = \sum_{j=1}^{n_i} Q_j \tag{9}$$
where $n_i$ is the rank of $Q_i$ and $\sum_i n_i = N$.

In fact, because $Q_i$ is Hermitian, the eigen-decomposition of $Q_i$ can be written as
$$Q_i = \sum_{j=1}^{n_i} \lambda_j |\psi_j\rangle\langle\psi_j| \tag{10}$$
where $|\psi_j\rangle$ is an elementary vector of the Hilbert space $H$. Therefore, we can always assume that
$$Q_j = \lambda_j |\psi_j\rangle\langle\psi_j| = |\mu_j\rangle\langle\mu_j| \tag{11}$$
where $|\mu_j\rangle = \sqrt{\lambda_j}|\psi_j\rangle$, is called the measurement vector.

Then our purpose is to find a set of corresponding PVM operators $\{P_j, i \in N\}$ for rank 1 POVM operators $\{Q_j, i \in N\}$. Suppose the dimension of U is $k \leq N$. Define a measurement matrix as a matrix consist of all measurement vectors:

$$M = \{|\mu_1\rangle, |\mu_2\rangle, \ldots, |\mu_N\rangle\} \tag{12}$$

According to the completeness condition of the POVM operators, we have
$$\sum_{j=1}^{N} Q_j = \sum_{j=1}^{N} |\mu_j\rangle\langle\mu_j| = MM^\dagger = P_u \tag{13}$$
where $P_u$ is a projector mapping from Hilbert space $H$ to subspace $U$.

The singular value decomposition of the measurement matrix is given by
$$M = U\Sigma V^\dagger \tag{14}$$
where $U$ and $V$ are unitary operators with dimension $n \times n$ and $N \times N$, respectively. Let $\{u_i, i = 1, \ldots, n\}$ be the column vectors of $U$ and $\{v_i, i = 1, \ldots, N\}$ be the column vectors of V. And $\Sigma$ is a $n \times N$ matrix in a form like [5]
$$\Sigma = \begin{vmatrix} I_k & 0 \\ 0 & 0 \end{vmatrix} \tag{15}$$
Therefore, the measurement matrix can be rewritten as
$$M = \sum_{i=1}^{k} |u_i\rangle\langle v_i| \tag{16}$$
And the projector $P_u$ can be rewritten as
$$P_u = MM^\dagger = \left(\sum_{i=1}^{k} |u_i\rangle\langle v_i|\right)\left(\sum_{i=1}^{k} |v_i\rangle\langle u_i|\right) = \sum_{i=1}^{k} |u_i\rangle\langle u_i| \tag{17}$$
Then we can define a $n \times N$ matrix $\widetilde{M}$ as
$$\widetilde{M} = \sum_{i=1}^{N} |u_i\rangle\langle v_i| \tag{18}$$
It is easy to verify that
$$P_u \widetilde{M} = \left(\sum_{i=1}^{k} |u_i\rangle\langle u_i|\right)\left(\sum_{i=1}^{N} |u_i\rangle\langle v_i|\right) = \sum_{i=1}^{k} |u_i\rangle\langle v_i| = M \tag{19}$$
And the columns of $\widetilde{M}$ are orthogonal:
$$\widetilde{M}^\dagger \widetilde{M} = \left(\sum_{i=1}^{N} |v_i\rangle\langle u_i|\right)\left(\sum_{i=1}^{N} |u_i\rangle\langle v_i|\right) = I_N \tag{20}$$
Therefore, we can rewrite $\widetilde{M}$ in the form of column vectors as:
$$\widetilde{M} = \{|p_1\rangle, |p_2\rangle, \ldots, |p_N\rangle\} \tag{21}$$
where $\{|p_i\rangle, i = 1,2,\ldots, N\}$ forms an orthogonal basis of the enlarged N-dimensional subspace $\widetilde{U}$. And we can define a set of projection operators $\{P_i, i = 1,2,\ldots, N\}$ as
$$P_i = |p_i\rangle\langle p_i| \tag{22}$$
Obviously, $\{P_i, i = 1,2,\ldots, N\}$ forms a PVM. And because $M = P_A \widetilde{M}$, i.e.
$$\{|\mu_1\rangle, |\mu_2\rangle, \ldots, |\mu_N\rangle\} = P_A\{|p_1\rangle, |p_2\rangle, \ldots, |p_N\rangle\} \tag{23}$$
Thus, for $i = 1,2,\ldots, N$:

$$|\mu_i\rangle = P_A |p_i\rangle; \quad Q_i = P_A P_i P_A \tag{24}$$
Therefore, if we perform a POVM measurement specified by $\{Q_i, i = 1,2,\ldots, N\}$ on an initial state $|\psi\rangle$ of subspace $U$, the probability of obtaining outcome $i$ is given by
$$p_n(\text{i}|\psi) = P(N = \text{i}|s = |\psi\rangle) = \langle\psi|Q_i|\psi\rangle = \langle\psi|P_A P_i P_A|\psi\rangle\rangle = \langle\psi|P_i|\psi\rangle \tag{25}$$
since $P_A|\psi\rangle = |\psi\rangle$. This means that the POVM specified by $\{Q_i, i = 1,2,\ldots, N\}$ in the original subspace U can be achieved by a PVM specified by $\{P_i, i = 1,2,\ldots, N\}$ in a lager subspace $\widetilde{U}$.

## 4. An example for POVM in practical measurement

Suppose that an initial quantum state $|\psi\rangle_U$ is a n-dimensional vector on 2-dimentional subspace U of a n-dimensional Hilbert space H, and we can denote it as:
$$|\psi\rangle_U = \alpha_0 |0\rangle_U + \alpha_1 |1\rangle_U \tag{26}$$
where $\{|0\rangle_U, |1\rangle_U\}$ is a basis of subspace U. and $|\alpha_0|^2 + |\alpha_1|^2 = 1$.

To perform a POVM on $|\psi\rangle_U$, we introduce an ancilla system which is independent from original system and choose an initial state $|0\rangle_{U^\perp}$, which is a n-dimensional vector on subspace $U^\perp$ of Hilbert space H, as an ancilla state. Then we can denote the combined state as $|\psi\rangle_U \otimes |0\rangle_{U^\perp}$, a state of the Hilbert space $H = U \otimes U^\perp$.

Then we perform a unitary operation A on the combined state to make them entangle with each other:
$$|\psi\rangle_U \otimes |0\rangle_{U^\perp} \to A(|\psi\rangle_U \otimes |0\rangle_{U^\perp}) \tag{27}$$

Then we perform a local PVM $\{P_i, i = 0, 1, ..., N-1\}$ on the ancilla system, which means we perform a PVM $\{\widehat{P}_i = I_U \otimes P_i, i = 0, 1, ..., N-1\}$ on the combined system. The probability of outcome $i$ is given by

$$p_i = (\langle\psi|_U \otimes \langle 0|_{U^\perp}) A^\dagger \widehat{P}_i A (|\psi\rangle_U \otimes |0\rangle_{U^\perp}) \tag{28}$$

Note that $\{O_i = A^\dagger \widehat{P}_i A, i = 0,1, ..., N-1\}$ is still PVM of the combined system $H$. We can represent the operator $O_i$ in the following expression:

$$O_i = \sum_{j,k} O_{jk}^{(i)} |j\rangle_U \langle j|_U \otimes |k\rangle_{U^\perp} \langle k|_{U^\perp} \tag{29}$$

where $\{|j\rangle_U, j = 0, 1\}$ is a basis of $U$ and $\{|k\rangle_{U^\perp}, k = 0, 1, ..., N-1\}$ is a basis of $U^\perp$. Substitute (26) and (29) into (28), we obtain:

$$p_i = \sum_j \alpha_j^* O_{j0}^{(i)} \alpha_j \tag{30}$$

Therefore, we can define an operator $Q_i$ as:

$$Q_i = \sum_j O_{j0}^{(i)} |j\rangle_U \langle j|_U \tag{31}$$

such that

$$p_i = \sum_j \alpha_j^* O_{j0}^{(i)} \alpha_j = \langle\psi|Q_i|\psi\rangle \tag{32}$$

In this case, the PVM measurement $\{O_i, i = 0, 1, ..., N-1\}$ of the combined system gives the same probabilities as the POVM measurement $\{Q_i, i = 0, 1, ..., N-1\}$ of the original system.

## 5. Conclusion

This paper is a brief introduction to POVM and Neumark's theorem. A concrete example of POVM measurement achieved by a PVM measurement is given. It provides some insight of POVM measurement of quantum communications.